\documentclass[letterpaper, 10 pt, conference]{ieeeconf}  

\IEEEoverridecommandlockouts                              
\usepackage{color}
\usepackage{arydshln}
\usepackage{booktabs} 
\usepackage{graphicx}
\usepackage{MnSymbol}
\usepackage[linesnumbered,ruled]{algorithm2e}
\usepackage{tikz}
\usepackage{tkz-euclide}
\usepackage{tikz-qtree}
\usepackage[utf8]{inputenc}

\newtheorem{proposition}{Proposition}
\newtheorem{problem}{Problem}
\newtheorem{definition}{Definition}
\newtheorem{remark}{Remark}

\newtheorem{example}{Example} 

\title{\LARGE \bf
Control from Signal Temporal Logic Specifications with \\ Smooth Cumulative Quantitative Semantics
} 

\author{Iman Haghighi, Noushin Mehdipour, Ezio Bartocci, Calin Belta
\thanks{This work was partially supported at Boston University by the NSF under grants IIS-1723995 and CMMI-1400167.} %
\thanks{Iman Haghighi, Noushin Mehdipour, and Calin Belta are with Boston University, Boston, MA 02215, USA {\tt\small \{haghighi,noushinm,cbelta\}@bu.edu}} %
\thanks{Ezio Bartocci is with TU Wien, Vienna, Austria {\tt\small ezio.bartocci@tuwien.ac.at}} %
}

\begin{document}

\maketitle
\thispagestyle{empty}
\pagestyle{empty}

\begin{abstract}
We present a framework to synthesize control policies for nonlinear dynamical systems from complex temporal constraints specified in a rich temporal logic called Signal Temporal Logic (STL).
We propose a novel smooth and differentiable STL quantitative semantics called cumulative robustness, and efficiently compute control policies through a series of smooth optimization problems that are solved using gradient ascent algorithms.
Furthermore, we demonstrate how these techniques can be incorporated in a model predictive control framework in order to synthesize control policies over long time horizons.
The advantages of combining the cumulative robustness function with smooth optimization methods as well as model predictive control are illustrated in case studies. 
\end{abstract}

\section{Introduction}
\label{sec:introduction}

In the last decade, formal methods have become powerful mathematical tools 
not only for specification and verification of systems, but also 
to enable control engineers to move beyond classical notions 
such as stability and safety, and to synthesize controllers 
that can satisfy much richer 
specifications~\cite{belta2017formal}. 
For example, temporal logics such as Linear Temporal 
Logic (LTL)~\cite{Pnueli77}, Metric Temporal 
Logic (MTL)~\cite{Koymans1990}, and Signal Temporal 
Logic (STL)~\cite{maler2004monitoring} have 
been used to define rich time-dependent constraints for control systems in a wide variety 
of applications, ranging from biological networks to multi-agent 
robotics~\cite{WongpiromsarnTM12,annpureddy2011s,raman2015reactive}.

The existing methods for temporal logic control can be divided into two general categories: 
automata-based \cite{belta2017formal} and 
optimization- based \cite{KaramanSF08,saha2016milp,pant2017smooth}. In the first, a finite abstraction for the system and an automaton representing the temporal logic specifications are computed.  A controller is then synthesized by solving a game over the product automaton~\cite{belta2017formal}.  Even though 
this approach has shown some promising results, automata-based 
solutions are generally very expensive computationally. The second approach leverages the definition of quantitative 
semantics~\cite{donze2010robust,rodionova2016temporal,akazaki2015time} 
for temporal logics that interpret a formula  with respect to a 
system trajectory by computing a real-value (called \textit{robustness}) 
measuring how strongly a specification is satisfied or violated.  
Consequently, the control problem becomes an optimization problem
 with the goal of maximizing robustness.

In this paper, we propose a novel framework to synthesize cost-optimal 
control policies for possibly nonlinear dynamical systems under STL constraints. 
First, we introduce a new quantitative semantics for STL, called 
\textit{cumulative} quantitative semantics or \textit{cumulative} robustness degree.
The traditional STL robustness degree~\cite{donze2010robust} is very conservative since it only considers the robustness at the most critical time (the time that is closest to violation). On the other hand, we cumulate the robustness over the time horizon of the specification.
This results in a robustness degree that is more useful and meaningful in 
many control applications.
Specifically, we show that control policies obtained by optimizing the robustness introduced here leads to reaching desired states faster, and the system spends more time in those states.

We demonstrate how to extend smooth approximation techniques described in~\cite{pant2017smooth} to address the novel notion of cumulative robustness introduced here.  We then show how to 
leverage the smooth cumulative robustness function to perform 
Model Predictive Control (MPC)~\cite{camacho2013model} 
under STL constraints for nonlinear dynamical systems using a gradient ascent algorithms.  We evaluate our approach on three different case studies: (1) path planning for an autonomous vehicle, (2) MPC for linear systems and (3) MPC for noisy linear systems. 

The rest of the paper is organized as follows. In Section~\ref{sec:related}, 
we discuss the related work.  In Section~\ref{sec:preliminaries}, we introduce 
the necessary theoretical background.
Section~\ref{sec:problem} presents the control problems considered in this paper. 
Section~\ref{sec:robustness} shows how to compute the cumulative robustness and its smooth approximation.  
Sections~\ref{sec:optimization} and \ref{sec:mpc} provide the optimization 
framework to perform model predictive control using gradient ascent algorithm for the 
smooth cumulative robustness function.
The capabilities of our algorithms are demonstrated through case studies in Section~\ref{sec:case}.
\section{Related Work and Contributions}
\label{sec:related}
In~\cite{akazaki2015time}, the authors introduce an extension of STL, called AvSTL, 
and propose a quantitative semantics for it. The AvSTL quantitative semantics has some similarities with the cumulative robustness proposed here, but computes the average robustness over specification horizons instead of cumulating the robustness. 
Moreover, the work in~\cite{akazaki2015time} only investigates a falsification problem, 
while we consider a more general control problem.

In~\cite{KaramanSF08}, Karaman et al. demonstrate that temporal logic control 
problems can be formulated as mixed integer linear problems (MILP), avoiding the issues with state space abstraction and dealing with systems in continuous space. 
Since this paper, many researchers have adopted this technique and demonstrated 
Mixed Integer Linear or Quadratic Programs (MILP/MIQP) are often more scalable 
and reliable than automata-based solutions~\cite{raman2014model,kim2017dynamic}.
However, MILP has an exponential complexity with respect to the number of its integer variables and the computational times for MILP-based solutions are extremely unpredictable. 
These types of solutions generally suffer when dealing with very large and nested specifications.
More recently, Pant et al. in~\cite{pant2017smooth} have presented a technique to compute a 
smooth abstraction for the traditional STL quantitative semantics defined in~\cite{donze2010robust}. 
The authors show that control problems can be solved using smooth optimization algorithms such as gradient descent in a much more time efficient way than MILP. This technique also works for any smooth nonlinear dynamics while MILP and MIQP require the system dynamics to be linear or quadratic. 
Additionally, the same smoothing technique enabled the authors in~\cite{li2017policy} to solve reinforcement learning problems.
We show here how to extend this
technique to smoothly approximate the proposed cumulative robustness function improving the control synthesis. 

We consider the problem of Model Predictive Control (MPC) under 
temporal logics constraints that is based on iterative, finite-horizon optimization 
of a plant model and its input in order to satisfy a signal temporal logic specification.
%
The work in~\cite{WongpiromsarnTM12} employs MPC to control 
the plant to satisfy a formal specification  expressed in a fragment of LTL that can be translated 
into finite state automata and~\cite{DingLB14} used it for finite state systems and B\"uchi 
automata to synthesize controllers under full LTL specification.
MPC has also been used in conjunction with mixed integer linear and quadratic programs for temporal logic control~\cite{raman2014model,BemporadM99,lindemann2019robust}. 

The methodology presented in this paper has some connections to economic model predictive control (EMPC)~\cite{ellis2014tutorial}, in that both frameworks consider cost functions which are more general and complex than the simple quadratic costs of conventional MPC. While EMPC defines costs over system state and input values, signal temporal logic control synthesis can be thought of as optimization problems with objective functions over system trajectories.
\section{preliminaries}
\label{sec:preliminaries}

\subsection{Signal Temporal Logic}
\label{sec:stl}

Signal Temporal Logic (STL) was introduced in \cite{maler2004monitoring}.
Consider a discrete unbounded time series $\tau:=\{t_k|k\in\mathbb{Z}_{\geq 0}\}$. A signal is a function $\sigma:\tau\rightarrow\mathbb{R}^n$ that maps each time point $t_k\in\mathbb{R}_{\geq 0}$ to an $n$-dimensional vector of real values $\sigma[k]$, with $\sigma _i[k]$ being the $i$th component.
Given a signal $\sigma$ and $k\in\mathbb{Z}_{\geq 0}$, $\sigma^{t_k}:=\{\sigma[k^\prime]|k^\prime\geq k\}$ is the portion of the signal starting at the $k$th time step of $\tau$.
\begin{definition}[STL Syntax]
\label{def:syntax}
\begin{equation}
\label{eq:syntax}
\varphi:=\top\; | \; \mu \; | \; \neg\varphi \; | \; \varphi_1\wedge\varphi_2 \;  | \; \varphi_1\mathbf{U}_I \varphi_2,
\end{equation}
\end{definition}
where $\phi$, $\varphi_1$, $\varphi_2$ are formulas and $\top$ stands for the Boolean constant True. We use the standard notation for the Boolean operators. $I=[k_1,k_2]$ denotes a bounded time interval containing all time points (integers) starting from $k_1$ up to $k_2$ and $k_2> k_1\geq 0$.
The building blocks of STL formulas are predicates of the form $\mu:=l(\sigma)\geq 0$ where $l$ is a linear or nonlinear combination of the elements of $\sigma$ (e.g., $\sigma_1+2\sigma_2-3\geq 0$ or $\sigma_1^2+\sigma_2^2-1 \geq 0$).
In this paper, we assume that $l$ is smooth and differentiable.
In order to construct a STL formula, different predicates ($\mu$) or STL sub-formulas ($\varphi$) are recursively combined using Boolean logical operators ($\neg,\vee,\wedge$) as well as temporal operators. 
$\mathbf{U}_I$ is the \textit{until} operator and $\varphi_1\mathbf{U}_I\varphi_2$ is interpreted as "$\varphi_2$ must become true at least once in the future within $I$ and $\varphi_1$ must be always true prior to that time".
Other temporal operator can also be derived from $\mathbf{U}_I$.
$\mathbf{F}_I$ is the \textit{finally} or \textit{eventually} operator and $\mathbf{F}_I\varphi$ is interpreted as "$\varphi$ must become true at least once in the future within $I$".
$\mathbf{G}_I$ is the \textit{globally} or \textit{always} operator and $\mathbf{G}_I\varphi$ is interpreted as "$\varphi$ must always be true during the interval $I$ in the future".
For instance, $\mathbf{G}_{[0,10]}(\sigma_1>0)$ means $\sigma_1$ must be positive all the time from now until $10$ units of time from now; $\mathbf{F}_{[0,5]}\mathbf{G}_{[0,10]}(\sigma_1>0)$ means that $\sigma_1$ must become positive within $5$ units of time in the future and stay positive for $10$ steps after that; and $(\sigma_2>0)\mathbf{U}_{[0,10]}(\sigma_1>0)$ means that $\sigma_1$ should become positive at a time point within $10$ units of time and $\sigma_2$ must be always positive before that.

STL has a qualitative semantics which can be used to determine whether a formula $\varphi$ with respect to a given signal $\sigma^{t_k}$ starting at the $k$th time step of $\tau$ is satisfied ($\sigma^{t_k}\models\varphi$) or violated ($\sigma^{t_k}\not\models\varphi$)   \cite{maler2004monitoring,donze2010robust}. 
Additionally, \cite{donze2010robust} introduces a quantitative semantics that can be interpreted as "How much a signal satisfies or violates a formula".
The quantitative valuation of a STL formula $\varphi$ with respect to a signal $\sigma$ at the $k$th time step is denoted by $\rho(\varphi,\sigma,t_k)$ and called the \textit{robustness degree}.
The robustness degree for any STL formula at time $t_k\in\tau$ with respect to signal $\sigma$ can be recursively computed using the following definition.

\begin{definition}[STL robustness]
\label{def:robustness}
\begin{equation}
\label{eq:stl}
\begin{array}{lcr}
\rho(\top,\sigma,t_k)&:=&+\infty,\\
\rho(l(\sigma)\geq 0,\sigma,t_k)&:=&l(\sigma[k]),\\
\rho(\neg\varphi,\sigma,t_k)&:=&-\rho(\varphi,\sigma,t_k),\\
\rho(\psi\wedge\varphi,\sigma,t_k)&:=&\min\{\rho(\psi,\sigma,t_k),\rho(\varphi,\sigma,t_k)\},\\
\rho(\psi\mathbf{U}_I\varphi,\sigma,t_k)&:=&\max\limits_{k^\prime\in I}(\min\{\rho(\varphi,\sigma,t_{k+k^\prime}),\\
&&\min\limits_{k^{\prime\prime}\in[k,k+k^\prime]}\rho(\psi,\sigma,t_{k^{\prime\prime}})\}).\\
\end{array}
\end{equation}
\end{definition}

The robustness degree for $\vee$, $\mathbf{F}_I$, and $\mathbf{G}_I$ can be easily derived~\cite{donze2010robust}. 
The robustness degree is sound, meaning that:
\begin{equation}
\label{eq:sound1}
\begin{array}{c}
\rho(\varphi,\sigma,t_k)>0\Rightarrow\sigma^{t_k}\models\varphi,\\
\rho(\varphi,\sigma,t_k)<0\Rightarrow\sigma^{t_k}\not\models\varphi.\\
\end{array}
\end{equation}

A formal definition for the \textit{horizon} ($h_\varphi$) of a STL formula $\varphi$ is presented in \cite{dokhanchi2014line}. Informally, it is the smallest time step in the future for which signal values are needed to compute the robustness for the current time point. For instance, the horizon of the formula $\mathbf{F}_{[0,5]}\mathbf{G}_{[0,10]}(\sigma_1>0)$ is $5+10=15$.


In the rest of this paper, if we do not specify the time of satisfaction or violation of a formula, we mean satisfaction or violation at time $0$ (i.e., $\sigma\models\varphi$ means $\sigma^0\models\varphi$).

\subsection{Smooth Approximation of STL Robustness Degree}
\label{sec:smooth}

The robustness degree that results from \eqref{eq:stl} is not differentiable. 
This poses a challenge in solving optimal control problems using the robustness degree as part of the objective function. 
The authors in \cite{pant2017smooth} introduce a technique for computing smooth approximations of the robustness degree for Metric Temporal Logic (MTL), which is based on smooth approximations of the max and min functions:


\begin{definition}[Smooth Operators]
\label{def:smooth}
\begin{equation}
\label{eq:smooth}
\begin{array}{l}
\widetilde{\max_\beta}(a_1,\ldots,a_m) := \frac{1}{\beta}\ln\sum_{i=1}^m e^{\beta a_i}, \\
\widetilde{\min_\beta}(a_1,\ldots,a_m):=-\widetilde{\max}(-a_1,\ldots,-a_m).
\end{array}
\end{equation}
\end{definition}
\vspace{5pt}
It is easily shown that \cite{pant2017smooth}:
\begin{equation}
\label{eq:smoothError}
\begin{array}{l}
0 \leq \widetilde{\max_\beta}(a_1,\ldots,a_m) - \max(a_1,\ldots,a_m) \leq \frac{\ln(m)}{\beta}, \\
0 \leq \widetilde{\min_\beta}(a_1,\ldots,a_m) - \min(a_1,\ldots,a_m) \leq \frac{\ln(m)}{\beta},
\end{array}
\end{equation}

Therefore, the approximation error approaches $0$ as $\beta$ goes to $\infty$.
We denote the smooth approximation of any robustness function $\rho$ by $\tilde{\rho}$.
\section{Problem Statement}
\label{sec:problem}

Consider a discrete time continuous space dynamical system of the following form:
\begin{equation}
\label{eq:dynamics}
\begin{array}{l}
\sigma[k+1]=f(\sigma[k],u[k]),\\
\sigma[0]=\gamma,\\
\end{array}
\end{equation}
where $\sigma[k]\in\mathcal{X}\subseteq\mathbb{R}^n$ is the state of the system at the $k$th time step of $\tau:=\{t_k|k\in\mathbb{Z}_{\geq 0}\}$, $\mathcal{X}$ is the state space, and $\gamma\in\mathcal{X}$ is the initial condition. 
$u[k]\in\mathcal{U}\subseteq\mathbb{R}^m$ is the control input at time step $k$ that belongs to a hyper-rectangle control space $\mathcal{U}=[\mathcal{U}_1,\mathcal{U}_1^\prime]\times\ldots\times[\mathcal{U}_m,\mathcal{U}_m^\prime]$.
$f$ is a smooth function representing the dynamics of the system.
The $i$th component of $\sigma$, $u$, $f$, and $\gamma$ are denoted by $\sigma_i$, $u_i$, $f_i$, and $\gamma_i$, respectively.
The system trajectory ($n$-dimensional signal) produced by applying control policy $u=\{u[k]\}$ is denoted by $\langle \sigma,u \rangle$.

A specification over the state of the system is given as a STL formula $\varphi$ with horizon $h_\varphi$.
We also consider a cost function $J:\mathcal{X}\times\mathcal{U}\rightarrow\mathbb{R}$ where $J(\sigma[k],u[k])$ is a smooth function representing the cost of applying the control input $u[k]$ at state $\sigma[k]$.
In the first problem, we intend to determine a control policy $u^*=\{u^*[k]|k=0,\ldots,h_\varphi-1\}$ over the time horizon of the specification $\varphi$ such that $\varphi$ is satisfied, while optimizing the cumulative cost.
\begin{problem}[Finite Horizon Control]
\label{problem1}
\begin{equation}
\label{eq:problem1}
\begin{array}{c}
u^*=\arg\min \sum\limits_{k=0}^{h_\varphi-1} J(\sigma[k],u[k]),\\
\text{s.t.}\;\;\;\;\; \langle\sigma,{u^*}\rangle\models\varphi.
\end{array}
\end{equation}
Furthermore, we intend to find the control policy that results in highest possible robustness degree.
\end{problem}

\begin{remark}
A solution to Problem \ref{problem1} based on mixed integer linear programming was presented in \cite{raman2014model}.
A smooth gradient descent solution was also provided in \cite{pant2017smooth}.
In both cases, the quantitative semantics defined in \eqref{eq:stl} was used. 
In this paper, we utilize an alternative quantitative semantics. 
We will demonstrate in Section \ref{sec:case} that this will result in a better control synthesis in certain types of applications.
\end{remark}

\begin{example}
\label{ex1}
\begin{figure}
\centering
\begin{tikzpicture}[
xscale=0.5,yscale=0.5
]
\draw[-,line width = 0.25mm,fill=cyan!10] (0.4,-3) -- (3,-3) -- (3,-1.3) -- (0.4,-1.3) -- cycle;
\draw[-,line width = 0.25mm,fill=cyan!10] (-3,0.4) -- (-1.3,0.4) -- (-1.3,3) -- (-3,3) -- cycle;
\draw[-,line width = 0.25mm,fill=green!10] (1.3,1.3) -- (3,1.3) -- (3,3) -- (1.3,3) -- cycle;
\draw[-,line width = 0.25mm,fill=red!10] (-1.3,-1.3) -- (1.3,-1.3) -- (1.3,1.3) -- (-1.3,1.3) -- cycle;
\draw[-,line width = 0.5mm] (-3,-3) -- (3,-3) -- (3,3) -- (-3,3) -- cycle;
\tkzText[](2,-2){1}
\tkzText[](-2,2){2}
\tkzText[](2.25,2.25){3}
\tkzText[](0,0){4}
\tkzText[below](3,-3){$7$}
\tkzText[left](-3,3){$7$}
\tkzText[below](0.4,-3){$4$}
\tkzText[left](-3,-1.3){$2$}
\tkzText[below](-1.3,-3){$2$}
\tkzText[left](-3,0.4){$4$}
\tkzText[below](1.3,-3){$5$}
\tkzText[left](-3,1.3){$5$}
\draw[line width=0.2mm, dashed,color=black!40]  (-1.3,-3) -- (-1.3 ,0.4);
\draw[line width=0.2mm, dashed,color=black!40]  (1.3,-3) -- (1.3 ,1.3);
\draw[line width=0.2mm, dashed,color=black!40]  (-3,-1.3) -- (0.4 ,-1.3);
\draw[line width=0.2mm, dashed,color=black!40]  (-3,1.3) -- (1.3 ,1.3);
\draw [->, line width=0.5mm] (-3,-3) -- (-2.5,-3);
\draw [->, line width=0.5mm] (-3,-3) -- (-3,-2.5);
\tkzText[below](-3,-3){$0$}
\tkzText[left](-3,-3){$0$}
\draw[fill=black!40,rotate around={45:(-2.1,-2.1)}] (-2.1,-2.1) ellipse (3mm and 1.5mm);
\draw (-2.1,-2.1) node[] {\textcolor{black} \textbullet};
\tkzText[below](-2.1,-3){$x$}
\tkzText[left](-3,-2.1){$y$}
\draw[line width=0.2mm, dashed,color=black!40]  (-2.1,-3) -- (-2.1 ,-2.1);
\draw[line width=0.2mm, dashed,color=black!40]  (-3,-2.1) -- (-2.1 ,-2.1);
\draw [->, line width=0.2mm] (-2.1,-2.1) -- (-1.5,-1.5);
\draw [->, dashed, line width=0.1mm] (-2.1,-2.1) -- (-1.5 ,-2.1);
\tkzText[](-1.65,-1.9){$\theta$}
\end{tikzpicture}
\caption{The workspace for the vehicle in Example \ref{ex1}. The goal is to visit regions 1 or 2 (cyan) in 3 steps and go to region 3 (green) in 7 steps after that and stay there for at least 2 steps, while avoiding region 4 (red).}
\label{fig:ex1}
\end{figure}
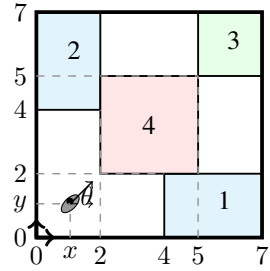
Consider an autonomous vehicle on a two dimensional square workspace (Fig. \ref{fig:ex1}).
The state of the vehicle consists of the horizontal and vertical position of its center as well as the heading angle ($\sigma=[x,y,\theta]$) The state space is $\mathcal{X}=[0,7]\times[0,7]\times\mathbb{R}$.
In Fig. \ref{fig:ex1}, the gray ellipse represents the vehicle.
The state evolves according to the following dynamics:
\begin{equation}
\label{eq:ex1dynamics}
\begin{array}{l}
x[k+1]=x[k]+\cos\theta[k]v[k]\Delta t,\\
y[k+1]=y[k]+\sin\theta[k]v[k] \Delta t,\\
\theta[k+1]=\theta[k]+v[k]\omega[k]\Delta t,\\
\end{array}
\end{equation}
where $u[k]=[v[k],\omega[k]]$ is the control input and belongs to the space $\mathcal{U}=[0,2]\times[-0.75,0.75]$.
$\Delta t$ is the time step size.
We consider the cost function $J(\sigma[k],u[k])=||\sigma[k+1]-\sigma[k]||_2^2$, assigning higher costs to motions over longer distances. 
Consider the following specification: "Eventually visit region 1 or 2 (cyan) within $6$ seconds. Afterwards, move to region 3 (green) in at most $4$ seconds and stay there for at least $2$ seconds, while always avoiding the unsafe region 4 (red)." Assuming $\Delta t=0.1$, this translates to:
\begin{equation}
\label{eq:ex1formula}
\phi_1=(\mathbf{G}_{[0,40]}\neg\mu_4)\mathbf{U}_{[0,60]}[(\mu_1\vee\mu_2)\wedge(\mathbf{F}_{[0,40]}\mathbf{G}_{[0,20]}\mu_3)],
\end{equation}
\vspace{-7pt}
where $\mu_i$ is the logical formula representing region $i$:
\begin{equation*}
\begin{array}{l}
\mu_1=x>4\wedge x<7 \wedge y>0\wedge y<2,\\
\mu_2=x>0\wedge x<2\wedge y>4\wedge y<7,\\
\mu_3=x>5\wedge x<7\wedge  y>5\wedge y<7,\\
\mu_4=x>2\wedge x<5\wedge y>2\wedge y<5.\\
\end{array}
\vspace{-4pt}
\end{equation*}
Note that the horizon of $\phi_1$ is $h_{\phi_1}=60+40+20=120$. 

We intend to determine the control policy $[v[k],\omega[k]]$ over this time horizon that satisfies this specification with highest possible robustness performance and minimum cost.
\end{example}

In the next problem, a time horizon $h_M$ is specified by the user such that $h_M\geq h_\varphi$ and we intend to find the control policy $u^*=\{u^*[k]|k=0,\ldots,h_M\}$ that results in the satisfaction of a given specification $\varphi$ at all times. 
\begin{problem}[Model Predictive Control for Finite Horizon]
\label{problem2}
\begin{equation}
\label{eq:problem2}
\begin{array}{c}
u^*=\arg\min \sum\limits_{k=0}^{h_\varphi+h_M-1}J(\sigma[k],u[k]),\\
\text{s.t.}\;\;\;\;\; \langle\sigma,u^*\rangle\models\mathbf{G}_{[0,h_M]}\varphi.\\
\end{array}
\end{equation}
\end{problem}
\vspace{5pt}
\begin{example}
\label{ex2}
Consider a linear system as follows:
\vspace{-5pt}
\begin{equation}
\label{eq:ex2dynamics}
x[k+1]=\begin{bmatrix}1&0.5\\0&0.8\end{bmatrix}x[k]+\begin{bmatrix}0\\1\end{bmatrix}u[k],
\end{equation}
where $x\in\mathbb{R}^2$, $x_1[0]=x_2[0]=0$, and $u[k]\in\mathbb{R}$. 
Consider the following STL specification.
\begin{equation}
\label{eq:ex2formula}
\phi_2=\mathbf{F}_{[0,4]}\mu_5\wedge\mathbf{F}_{[0,4]}\mu_6,
\end{equation}
where $\mu_5=(x_1>2\wedge x_1<4)$ and $\mu_6=(x_1<-2\wedge x_1>-4)$. $\phi_2$ requires the value of $x_1$ (the first component in $x$) to satisfy both $\mu_5$ and $\mu_6$ within $4$ time steps in the future.
Note that in this example, $h_{\phi_2}=4$.
Globally satisfying this specification for $h_M=15$ time steps ($\mathbf{G}_{[0,15]}\phi_2$) means that we require $x_1$ to periodically alternate between $\mu_5$ and $\mu_6$.  
\end{example}
\section{Smooth Cumulative Robustness}
\label{sec:robustness}

The robustness degree from Definition \ref{def:robustness} has been widely used in the past few years to solve control problems in various applications. 
Its soundness and correctness properties have enabled researchers to reduce complex control problems to manageable optimization problems. 
However, this definition is very conservative, since it only considers the system performance at the most critical time. 
Hence, any information about the performance of the system at other times is lost.
Inspired by \cite{akazaki2015time}, we introduce an alternative approach to compute the robustness degree, which we call the \textit{cumulative robustness}. 
This robustness is less conservative than \eqref{eq:stl}, and generally results in better performance if employed to solve control problems such as \eqref{eq:problem1} (see Section \ref{sec:case}).

For any STL formula $\varphi$, we define a \textit{positive cumulative robustness} $\rho^+(\varphi,\sigma,t_k)\in\mathbb{R}_{\geq 0}$ and a \textit{negative cumulative robustness} $\rho^-(\varphi,\sigma,t_k)\in\mathbb{R}_{\leq 0}$. 
For this purpose, we use two functions $\mathfrak{R}^+:\mathbb{R}\rightarrow\mathbb{R}_{\geq 0}$ and $\mathfrak{R}^-:\mathbb{R}\rightarrow\mathbb{R}_{\leq 0}$, which we call the positive and negative rectifier, respectively.
\begin{definition}[Rectifier Function]
\label{def:rectifier}
\begin{equation}
\label{eq:rectifier1}
\begin{array}{l}
\mathfrak{R}^+(a)=\max(0,a),\\
\mathfrak{R}^-(a)=\min(0,a).\\
\end{array}
\end{equation}
\end{definition}
\vspace{5pt}
We can use \eqref{eq:smooth} to smoothly approximate both rectifiers:
\begin{equation}
\label{eq:rectifier2}
\begin{array}{l}
\widetilde{\mathfrak{R}}_\beta^+(a)=\frac{1}{\beta}\ln(1+e^{\beta a}),\\
\widetilde{\mathfrak{R}}_\beta^-(a)=-\frac{1}{\beta}\ln(1+e^{-\beta a}).\\
\end{array}
\end{equation}
The positive and negative cumulative robustness are recursively defined as follows:

\begin{definition}[Cumulative Robustness]
\label{def:cumulative}
\begin{equation*}
\begin{array}{lcr}  
\rho^+(l(\sigma)\geq 0,\sigma,t_k)&:=&\mathfrak{R}^+(l(\sigma[k])),\\
\rho^-(l(\sigma)\geq 0,\sigma,t_k)&:=&\mathfrak{R}^-(l(\sigma[k])),\\
\rho^+(\neg\varphi,\sigma,t_k)&:=&-\rho^-(\varphi,\sigma,t_k),\\
\rho^-(\neg\varphi,\sigma,t_k)&:=&-\rho^+(\varphi,\sigma,t_k),\\
\rho^+(\psi\vee\varphi,\sigma,t_k)&:=&\max\{\rho^+(\psi,\sigma,t_k),\rho^+(\varphi,\sigma,t_k)\},\\
\rho^-(\psi\vee\varphi,\sigma,t_k)&:=&\max\{\rho^-(\psi,\sigma,t_k),\rho^-(\varphi,\sigma,t_k)\},\\
\rho^+(\psi\wedge\varphi,\sigma,t_k)&:=&\min\{\rho^+(\psi,\sigma,t_k),\rho^+(\varphi,\sigma,t_k)\},\\
\rho^-(\psi\wedge\varphi,\sigma,t_k)&:=&\min\{\rho^-(\psi,\sigma,t_k),\rho^-(\varphi,\sigma,t_k)\},\\
\rho^+(\mathbf{F}_I\varphi,\sigma,t_k)&:=&\sum\limits_{k^\prime\in I}\rho^+(\varphi,\sigma,t_{k+k^\prime}),\\
\rho^-(\mathbf{F}_I\varphi,\sigma,t_k)&:=&\sum\limits_{k^\prime\in I}\rho^-(\varphi,\sigma,t_{k+k^\prime}),\\
\rho^+(\mathbf{G}_I\varphi,\sigma,t_k)&:=&\min\limits_{k^\prime\in I}\rho^+(\varphi,\sigma,t_{k+k^\prime}),\\
\rho^-(\mathbf{G}_I\varphi,\sigma,t_k)&:=&\min\limits_{k^\prime\in I}\rho^-(\varphi,\sigma,t_{k+k^\prime}),\\
\end{array}
\end{equation*}
\begin{equation}
\label{eq:cumulative}
\begin{array}{lcr}
\rho^+(\psi\mathbf{U}_I\varphi,\sigma,t_k)&:=&\sum\limits_{k^\prime\in I}(\min\{\rho^+(\varphi,\sigma,t_{k+k^\prime}),\\
&&\min\limits_{k^{\prime\prime}\in[k,k+k^\prime]}\rho^+(\psi,\sigma,t_{k^{\prime\prime}})\}),\\
\rho^-(\psi\mathbf{U}_I\varphi,\sigma,t_k)&:=&\sum\limits_{k^\prime\in I}(\min\{\rho^-(\varphi,\sigma,t_{k+k^\prime}),\\
&&\min\limits_{k^{\prime\prime}\in[k,k+k^\prime]}\rho^-(\psi,\sigma,t_{k^{\prime\prime}})\}).\\
\end{array}
\end{equation}
\end{definition}

An extension to STL, called AvSTL, was introduced in \cite{akazaki2015time}. The authors employed AvSTL to solve falsification problems and did not consider the general control problem in that work.
The cumulative robustness for STL as defined in Definition \ref{def:cumulative} has two main differences from AvSTL.
First, we consider signals in discrete time.
Second, AvSTL computes the average robustness of $\mathbf{F}_I$ and $\mathbf{U}_I$ over their time intervals, while we cumulate the robustness.
Our purpose is to reward trajectories that satisfy the specification in front of $\mathbf{F}_I$ and $\mathbf{U}_I$ for longer time periods.

The positive cumulative robustness can be interpreted as robustness for portions of the signal that satisfy the formula and the negative cumulative robustness can be interpreted as robustness for portions of the signal that violate the formula. 
Our motivation for defining this alternative robustness degree was to modify the robustness of the \textit{finally} operator in order to cumulate the robustness for all the times in which the formula is true, whereas the traditional robustness degree only considers the most critical time point and does not take other portions of the signal into account.
However, this cannot be done in one robustness degree, since the positive and negative values of robustness cancel each other in time.
We divide the robustness into two separate positive and negative values to avoid this issue.

The following example demonstrates the advantages of the cumulative robustness \eqref{eq:cumulative} in comparison with \eqref{eq:stl}.

\begin{figure}
\centering
\includegraphics[width=0.65\columnwidth]{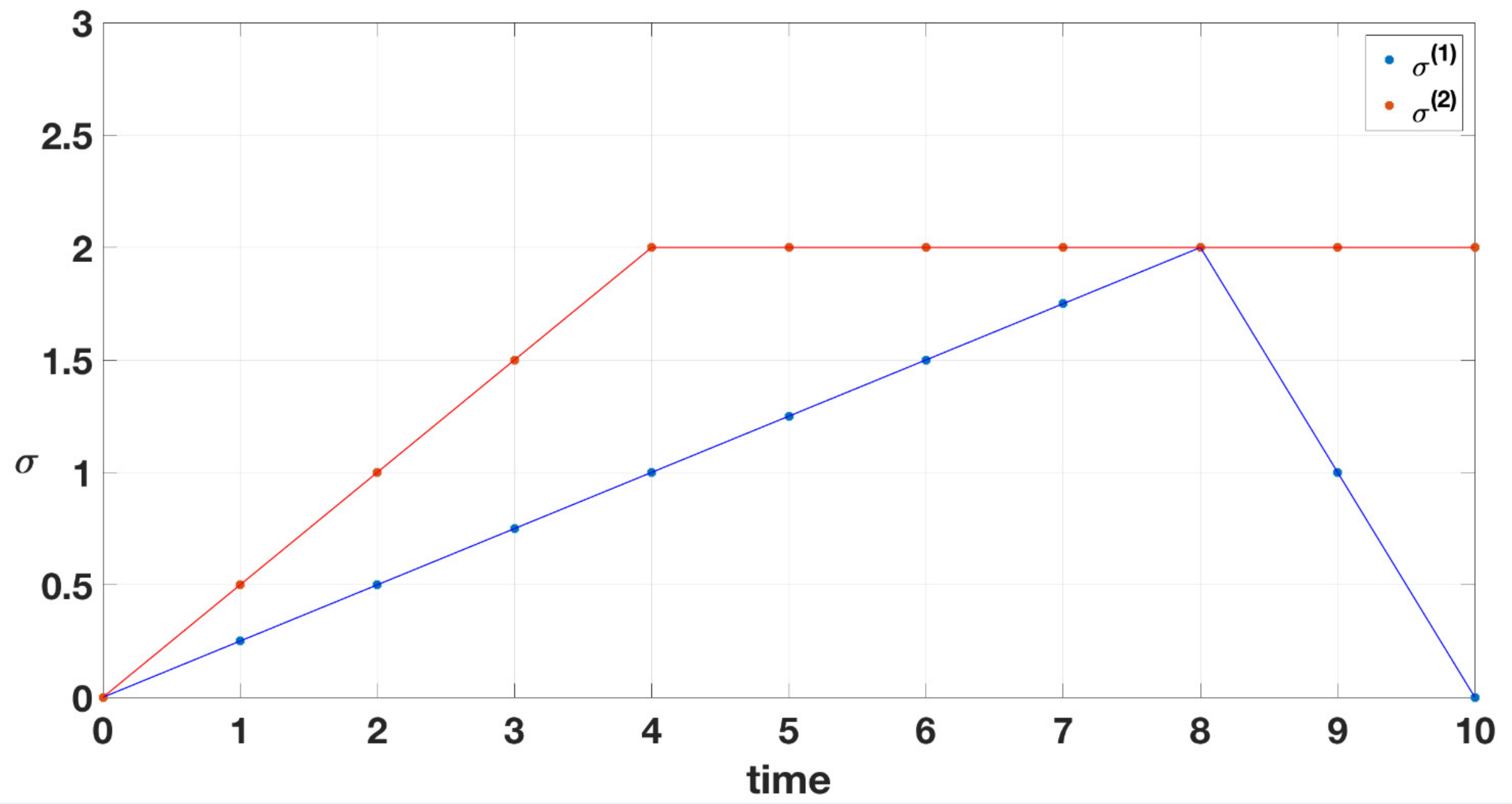}
\caption{Two discrete time signals that satisfy $\varphi_e=\mathbf{F}_{[0,10]}(\sigma>1\wedge \sigma<3)$ with similar robustness $\rho(\varphi,\sigma,0)$ but different positive cumulative robustness $\rho^+(\varphi,\sigma,0)$}.
\label{fig:ex3}
\end{figure}

\begin{example}
\label{ex3}
Consider the STL formula $\varphi_e=\mathbf{F}_{[0,10]}(\sigma>1\wedge \sigma<3)$ over a one dimensional signal $\sigma$ in a discrete time space $\tau=\{0,1,2,\ldots\}$. 
Fig. \ref{fig:ex3} demonstrates two different instances of this signal $\sigma^{(1)}$ and $\sigma^{(2)}$ starting at $\sigma[0]=0$, both satisfying the formula.
$\varphi_e$ has the same robustness score with respect to $\sigma^{(1)}$ and $\sigma^{(2)}$ at time 0, $\rho(\varphi_e,\sigma^{(1)},0)=\rho(\varphi_e,\sigma^{(2)},0)=1$. 
This is because the traditional robustness degree of the finally operator $\mathbf{F}_I\psi$ only returns the robustness of $\psi$ at the most critical time point in $I$, which is the same for both instances in this case.
However, the positive cumulative robustness of $\mathbf{F}_I\psi$ adds the robustness at all times in which $\psi$ is satisfied, hence rewarding signals that reach the condition specified by $\psi$ faster and stay in the satisfactory region longer. 
In this example, $\rho^+(\varphi_e,\sigma^{(1)},0)=2.5$ while $\rho^+(\varphi_e,\sigma^{(2)},0)=7.5$.
This is a desirable property in many control applications since by synthesizing controls that optimize the cumulative robustness for the finally operator, we are producing a trajectory of the system that reaches the specified condition as soon as possible and holds it true for as long as possible.
\end{example}

\label{prop:sound}
By comparing Definition \ref{def:cumulative} with Definition \ref{def:robustness}, it is easy to see that a STL formula is satisfied if the corresponding positive robustness is strictly positive.
\begin{proposition}
\begin{equation}
\label{eq:sound2}
\rho^+(\varphi,\sigma,t_k)>0\Leftrightarrow\rho(\varphi,\sigma,t_k)>0\Rightarrow\sigma^{t_k}\models\varphi.
\end{equation}
\end{proposition}
\vspace{5pt}
\label{prop:demorgan}
Furthermore, by combining different operators according to Definition \ref{def:cumulative}, it can be proven that:
\begin{proposition}
\begin{equation}
\label{eq:demorgan}
\begin{array}{l}
\rho^+(\neg(\neg\varphi),\sigma,t_k)=\rho^+(\varphi,\sigma,t_k),\\
\rho^+(\neg(\psi\wedge\varphi),\sigma,t_k)=\rho^+(\neg\psi\vee\neg\varphi,\sigma,t_k).\\
\end{array}
\end{equation}
The same properties hold for $\rho^-$ as well. This is important to consider since some of the similar robustness notions, such as the one introduced in \cite{rodionova2016temporal}, neglect these basic properties.
\end{proposition}

\begin{remark}
The causes of non-smoothness in \eqref{eq:cumulative} are the $\max$, $\min$, $\mathfrak{R}^+$, and $\mathfrak{R}^-$ functions.
Therefore, any cumulative robustness function can be smoothly approximated by replacing any appearance of these terms with their corresponding smooth approximations from \eqref{eq:smooth} and \eqref{eq:rectifier2}. 
The smooth approximation of $\rho^+$ and $\rho^-$ are denoted by $\tilde{\rho}^+$ and $\tilde{\rho}^-$.
\end{remark}

\begin{remark}
Cumulative robustness prohibits the \textit{globally} operator to be defined from \textit{finally}, meaning that $\rho^+(\mathbf{G}_I\varphi,\sigma,t_k)$ is not the same as $\rho^+(\neg\mathbf{F}_I\neg\varphi,\sigma,t_k)$. This is because we are altering the interpretation of \textit{finally} to mean "as soon as possible".
In the rest of this paper, we assume that any given STL formula $\varphi$ does not include $\neg\mathbf{F}_I{\varphi}$. Any such requirement can be specified with $\mathbf{G}_I\neg\varphi$ instead.
\end{remark}

\section{Smooth Optimization}
\label{sec:optimization}
In previous sections, we demonstrated that two different sound quantitative semantics (traditional robustness $\rho$ and positive cumulative robustness $\rho^+$) may be defined for STL specifications. 
Furthermore, they can be approximated as smooth functions, denoted by $\tilde{\rho}$ and $\tilde\rho^+$, respectively.
In this section, our approach to solve Problem \ref{problem1} is explained.

To ensure that the state $\sigma$ always remains in the state space $\mathcal{X}$, we add the following requirement to specification $\varphi$:
\begin{equation}
\label{eq:boundaries}
\varphi\gets\varphi\wedge \mathbf{G}_{[0,h_\varphi]}(\sigma\in\mathcal{X}).
\end{equation}

We aim to use $\tilde{\rho}^+$ in a gradient ascent setting similar to \cite{pant2017smooth} to solve this problem. 
However, note that $\rho^+=0$ any time that a formula is violated. 
Therefore, $\nabla\rho^+=0$ when $\rho^+=0$ unless the formula is on the boundaries of violation and very close to being satisfied. 
As a result, $\tilde\rho^+$ is almost guaranteed to fall in its local minimum when one initializes the gradient ascent algorithm for $\tilde\rho^+$, unless there is significant a priori knowledge about the system.
The negative cumulative robustness $\rho^-$ is not helpful either since there is no soundness theorem associated with it (see Section \ref{sec:robustness}).

We can circumvent this setback if we initialize the gradient ascent algorithm such that the specification is already satisfied and we only intend to maximize the level of satisfaction and minimize the cost.
Hence, we propose the three-stage algorithm $\textsc{SmoothOptimization}$ (Algorithm \ref{alg:optimization}).
The first stage aims to find a control policy that minimally satisfies $\varphi$, the second stage aims to maximize $\tilde\rho^+$, and the third stage aims to minimize the cumulative cost.
In each stage, one optimization problem is solved using the \textsc{GradientAscent} subroutine (Algorithm \ref{alg:gradient}).

Given a generic smooth objective function $Q(\sigma,u)$, an initial control policy $u^\iota=\{u^\iota[k]|k=1,\ldots,h_\varphi\}$, and a termination condition $\mathcal{T}$, Algorithm \ref{alg:gradient} uses gradient ascent \cite{bertsekas1999nonlinear} to find the optimal control policy $u^\Delta$. 
It starts by initializing a control policy $u^\iota$ for every time step $k\in\{0,\ldots,h_\varphi\}$, and computing the corresponding system trajectory $\langle\sigma,u\rangle$ from the system dynamics \eqref{eq:dynamics} (lines \ref{line:init1} to \ref{line:init2}).
At each gradient ascent iteration $i$, it updates the control policy for every time step according to the following rule (line \ref{line:update}).

\begin{equation}
\label{eq:gradient1}
u\gets u+\alpha_i\nabla Q,
\end{equation}

where $\alpha_i>0$. In the case studies presented in this paper, we used decreasing gradient coefficients $\alpha_{i+1}<\alpha_i$. However, one can employ other strategies that might work more efficiently depending on the application~\cite{bertsekas1999nonlinear}. Each component of $\nabla Q$ can be computed as (lines \ref{line:gradient1} to \ref{line:gradient2}):
\begin{equation}
\label{eq:partial}
\nabla_p Q[k]=\frac{\partial Q}{\partial u_p[k]}+\sum\limits_{q=1}^{n}\big(\frac{\partial Q}{\partial \sigma_{q}[k+1]}.\frac{\partial f_q[k+1]}{\partial u_p[k]}\big),
\end{equation}
for $p=1,\ldots,m$ and $k=0,\ldots,h_\varphi-1$. We continue this process until we find a control policy $u^\Delta$ that satisfies $\mathcal{T}$.

Recall that the problem setup in Section \ref{sec:problem} included constraints for both state ($\sigma[k]\in\mathcal{X}$) and control inputs ($u[k]\in\mathcal{U}$). we have already dealt with state constraints as part of the specification \eqref{eq:boundaries}.
Considering the fact that $\mathcal{U}=[\mathcal{U}_1,\mathcal{U}_1^\prime]\times\ldots\times[\mathcal{U}_m,\mathcal{U}_m^\prime]$ is assumed to be a hyper-rectangle, the input constraints are included using a projected gradient (line \ref{line:projection})~\cite{richter2009real}.

\begin{equation}
\label{eq:projection}
u_p[k]\gets\max\{\min\{u_p[k],\mathcal{U}^\prime_p\},\mathcal{U}_p\}.
\end{equation}
At any point in the optimization process, if a component of the control input $u_p[k]$ falls outside of the admissible interval $[\mathcal{U}_p,\mathcal{U}^\prime_p]$, we simply project it into the interval.

\begin{algorithm}
\caption{\textsc{SmoothOptimization}}
\label{alg:optimization}
\KwIn{STL Formula $\varphi$; Smooth Robustness Function $\tilde\rho(\varphi,\langle\sigma,u\rangle,0)$; Smooth Positive Cumulative Robustness Function $\tilde\rho^+(\varphi,\langle\sigma,u\rangle,0)$; Cost Function $J(\sigma[k],u[k])$; Initial Control Policy $u^0$; Smooth $m$-input $n$-output System Dynamics $f(\sigma[k],u[k])$; Initial State $\gamma$; Coefficients $\alpha_i$}
\KwOut{Optimal Control Policy $u^*$}
$u^{\Delta_1}\gets\textsc{GradientAscent}(\tilde\rho;u^0;\tilde\rho>0;f;\gamma;\alpha_i)$\;
$u^{\Delta_2}\gets\textsc{GradientAscent}(\tilde\rho^+;u^{\Delta_1};\nabla\tilde\rho^+<\epsilon;f;\gamma;\alpha_i)$\;
$u^*\gets\textsc{GradientAscent}(-\sum_k J;u^{\Delta_2};\tilde\rho^+<\xi;f;\gamma;\alpha_i)$\;
\end{algorithm}

In the first stage of \textsc{SmoothOptimization} (Algorithm \ref{alg:optimization}), we randomly initialize a control policy $u^0=\{u^0[k]|k=0,\ldots,h_\varphi-1\}$ and use the smooth approximation for traditional robustness function $\tilde{\rho}$ as the objective function in \textsc{GradientAscent}, only to find a control policy that minimally satisfies the specification ($\mathcal{T}$ is $\tilde\rho>0$).

In the second stage, we use the policy that we get at stage 1 to initialize a second \textsc{GradientAscent} with the objective function $\tilde\rho^+$ and termination condition $\nabla\rho^+<\epsilon$ where $\epsilon$ is a small positive number.
Note that $\tilde\rho^+$ is guaranteed to be strictly positive at the first iteration ($i=0$) due to \eqref{eq:sound2}.
Therefore, it will remain strictly positive as we proceed in the gradient ascent algorithm since this algorithm is designed such that the objective function only increases at each iteration \cite{bertsekas1999nonlinear}.
This prohibits the algorithm to ever fall in a local minimum at $\tilde\rho^+=0$.  

At the end of the second stage, we have a control policy $u^{\Delta_2}$ with a maximal level of cumulative robustness. 
However, we intend to minimize the cumulative cost $\sum_k J$ in Problem \ref{problem1}. 
We perform a third \textsc{GradientAscnet} with the objective function $-\sum_k J$ and control policy initialized at $u^{\Delta_2}$. 
In other words, we are updating the entire control policy gradually in order to decrease the cumulative cost at the expense of cumulative robustness.
This stage must terminate before the specification is violated (i.e., while the cumulative robustness is still strictly positive).
Hence, the termination condition at this stage is $\tilde\rho^+<\xi$ with $\xi>0$.

A small choice for $\xi$ results in a control policy with an almost optimal cost that minimally satisfies $\varphi$, while a large choice for $\xi$ results in a control policy with a higher level of cumulative robustness that is sub-optimal with respect to the cost.
The user can tune $\xi$ to achieve the desirable balance between the level of satisfaction and cost-optimality.

It is common in the literature to combine the robustness degree and control cost in a single optimization problem, using $\xi\tilde\rho-\sum J$ as the objective function \cite{pant2018tech}. 
We cannot do the same here since the combination of cumulative robustness $\tilde\rho^+$ with cost $\sum J$ in a single objective function can lead the value of $\tilde\rho^+$ to become zero again in the second stage of Algorithm \ref{alg:optimization}.
That is why we need to optimize cumulative robustness and cost in two separate stages.

\begin{algorithm}
\caption{\textsc{GradientAscent}}
\label{alg:gradient}
\KwIn{Smooth Objective Function $Q(\sigma,u)$; Initial Control Policy $u^\iota$; Termination Condition $\mathcal{T}$; Smooth $m$-input $n$-output System Dynamics $f(\sigma[k],u[k])$; Initial State $\gamma$; Coefficients $\alpha_i$}
\KwOut{Control Policy $u^{\Delta}$}
$u\gets u^\iota$\label{line:init1}\;
$\sigma[0]\gets \gamma$\;
\For{$k\gets 1,2,\ldots,h_\varphi$}{$\sigma[k+1]\gets f(\sigma[k],u[k])$\;}\label{line:init2}
$i\gets 0$\;
\While{$\neg\mathcal{T}$}{
\For{$p\gets 1,2,\ldots,n$
\label{line:gradient1}}{
$\delta_p[h_\varphi]\gets\frac{\partial Q}{\partial\sigma_p[h_\varphi]}$\;}
\For{$k\gets h_\varphi-1,h_\varphi-2,\ldots,1$}{
\For{$p\gets 1,2,\ldots,n$}{
$\delta_p[k]\gets\frac{\partial Q}{\partial\sigma_p[k]}+\sum\limits_{q=1}^n\big(\delta_q[k+1].\frac{\partial f_q[k+1]}{\partial \sigma_p[k]}$\big)\;}}
\For{$k\gets 0,1,\ldots,h_\varphi-1$}{
\For{$p\gets 1,2,\ldots,m$}{
$\zeta_p[k]\gets\frac{\partial Q}{\partial u_p[k]}+\sum\limits_{q=1}^n\big(\delta_q[k+1].\frac{\partial f_q[k+1]}{\partial u_p[k]}\big)$\label{line:gradient2}\;
$u_p[k]\gets u_p[k]+\alpha_i\zeta_p[k]$\label{line:update}\;
$u_p[k]\gets\max\{\min\{u_p[k],\mathcal{U}^\prime_p\},\mathcal{U}_p\}$\label{line:projection}\;}}
\For{$k\gets 1,2,\ldots,h_\varphi$}{$\sigma[k+1]\gets f(\sigma[k],u[k])$\;}
$i\gets i+1$\;}
\Return{$u$}
\end{algorithm}


\begin{remark}
If stage 1 does not terminate, it means that $\varphi$ is infeasible and we do not need to proceed to stage 2.
\end{remark}
\section{Model Predictive Control}
\label{sec:mpc}
In this section, we describe our solution to Problem \ref{problem2}.
Note that while Problem \ref{problem1} requires specification $\varphi$ to be satisfied at time $0$ (i.e., $\langle\sigma,u^*\rangle^0\models\varphi$), Problem \ref{problem2} requires it to be satisfied at all times in a $h_M$ horizon (i.e., $\langle\sigma,u^*\rangle^{t_k}\models\varphi\;\forall k\in\{0,\ldots,h_M\}$ or equivalently $\langle\sigma,u^*\rangle^0\models\mathbf{G}_{[0,h_M]}\varphi$).

The following procedure is performed to solve this problem. 
For time step $k=0$, we fix $\sigma[0]$ and use Algorithm \ref{alg:optimization} to synthesize control policy $u^{h_0}=\{u^{h_0}[k^\prime]|k^\prime=0,\ldots,h_\varphi-1\}$.
This ensures that $\phi$ is satisfied at time $t_0$.
Now, we only execute $u^{h_0}[0]$.
At time $k=1$, we fix $\sigma[1]$, use Algorithm \ref{alg:optimization} to synthesize $u^{h_1}=\{u^{h_1}[k^\prime]|k^\prime=0,\ldots,h_\varphi-1\}$, and only execute $u^{h_1}[0]$, which ensure satisfaction of $\varphi$ at time $t_1$.
We continue this process until we reach $k=h_M$.
Consequently, $\mathbf{G}_{[0,h_M]}\varphi$ is guaranteed to be satisfied for the following control policy, assuming that it is feasible at all times.

\begin{equation}
\label{eq:mpc}
u^*[k]=u^{h_k}[0]\;\;\;k=0,\ldots h_M.
\end{equation}
The smooth optimization procedure becomes particularly advantageous when employed instead of mixed integer programming in a model predictive control setting, since a separate optimization problem needs to be solved at every time step and smooth optimization has a much greater potential for being fast enough to be applied online.
Moreover, MILP solvers are very sensitive to the changes in the initial state $\gamma$, which is challenging since one has to solve a new MILP from scratch for any changes that occur in $\gamma$. This becomes a much more complex challenge when a MILP-based approach is used for multi-agent systems \cite{haghighi2016robotic}. On the other hand, smooth gradient ascent is much less sensitive to small changes in individual variables such as initial state $\gamma$ \cite{bertsekas1999nonlinear}.

Similar to \cite{raman2014model}, we can stitch together trajectories of length $h_M$ using a receding horizon approach to produce trajectories that satisfy $\mathbf{G}_{[0,\infty)}\varphi$. However, this does not guarantee recursive feasibility. In other words, we need to make sure that the resulting trajectory from the optimal controller consists of a loop. If this is the case, we can terminate the computation and keep repeating the control loop forever, which guarantees the satisfaction of the specification at all times, since we know that each loop satisfies the specification.

Formally, we add the following constraint to our optimization problem at every step.
\vspace{-5pt}
\begin{equation}
\label{eq:feasiblity}
\exists k\in\tau,\exists K\in\mathbb{N}\;\;s.t.\;\;\sigma[k+K]=\sigma[k],K>h_\varphi.
\end{equation}
This ensures that the resulting system trajectory contains a loop. If we find a solution to the optimization problem with positive robustness ($\rho>0$), this loop satisfies the given specification $\varphi$. Therefore, repeating the control strategy that produced this loop forever results in the satisfaction of the formula at all times, ensuring recursive feasibility. 

We use a brute force approach to find $k,K$ that satisfy \eqref{eq:feasiblity}. 
In other words, we start by guessing values for these parameters and keep changing them until we find values for which a solution to the optimization problem exists.
Fixing the values of $k$ and $K$ eliminates the quantifiers in \eqref{eq:feasiblity} and turns it into a linear constraint which is handled easily by any gradient descent solver through projection methods.

In the future, we plan to investigate other approaches to solve the problem with this additional constraint. 
\section{Case Study}
\label{sec:case}

In this section, we illustrate how our algorithms are able to solve Examples \ref{ex1} and \ref{ex2} in Section \ref{sec:problem}. All implementations were performed using MATLAB on a MacBook Pro with a 3.3 GHz Intel Core i7 processor and 16 GB of RAM.

\subsection{Example \ref{ex1}: Path Planning for Autonomous Vehicles}

\begin{figure}[t]
\centering
\includegraphics[width=\columnwidth]{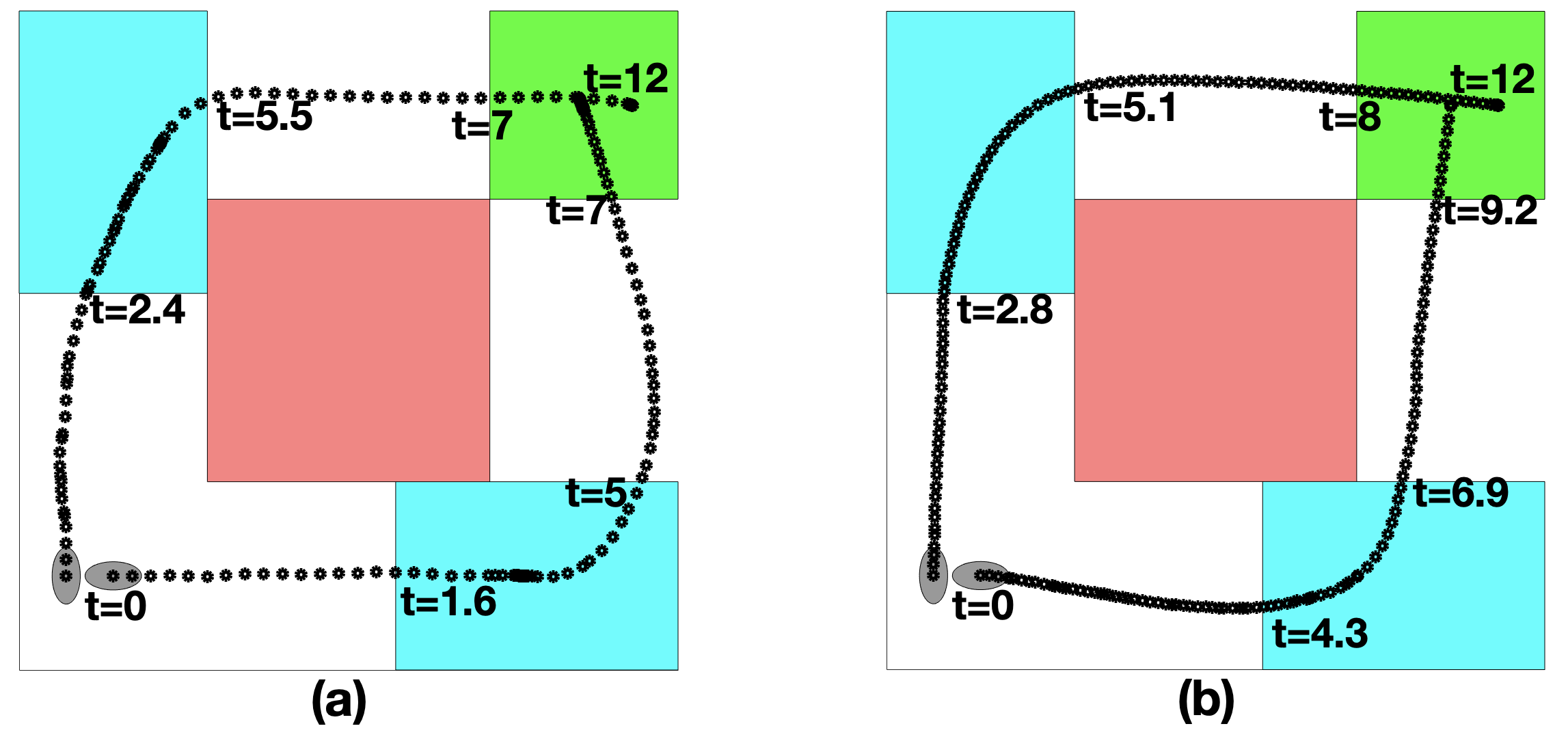}
\caption{The optimal path for two vehicles in Example \ref{ex1} with: (a) maximal cumulative robustness $\tilde\rho^+$, (b) maximal traditional robustness $\tilde\rho$. (The numbers next to each path indicate time.) }
\label{fig:ex1resultsNew}
\end{figure}

Consider the system and specification of Example \ref{ex1} with the dynamics of \eqref{eq:ex1dynamics} and specification $\phi_1$ presented in \eqref{eq:ex1formula}.
We assume that there are two vehicles (Yielding a 6 dimensional state space). Each vehicle starts from a different initial position and orientation, but follows the same dynamics. Both are required to follow the specification $\phi_1$ while avoiding collision. 
Recall that the objective is to visit one of the cyan regions in $6$ seconds and go to the green destination in at most $4$ seconds after that, while avoiding the red region (Fig. \ref{fig:ex1}). 
We used Algorithm \ref{alg:optimization} to solve this problem. The solution was computed in 18.3 seconds and the optimal path for each vehicle is shown in Fig. \ref{fig:ex1resultsNew}(a). The optimal path with respect to the traditional robustness score was also computed in 13.4 seconds and demonstrated in Fig. \ref{fig:ex1resultsNew}(b). It is obvious from these figures that optimizing the cumulative robustness results in paths in which the vehicles get to the their respective destinations faster and remain there longer. 

\subsection{Example \ref{ex2}: MPC for a Linear System}
\label{sec:case2}

\begin{figure}[t]
\centering
\includegraphics[width=0.8\columnwidth]{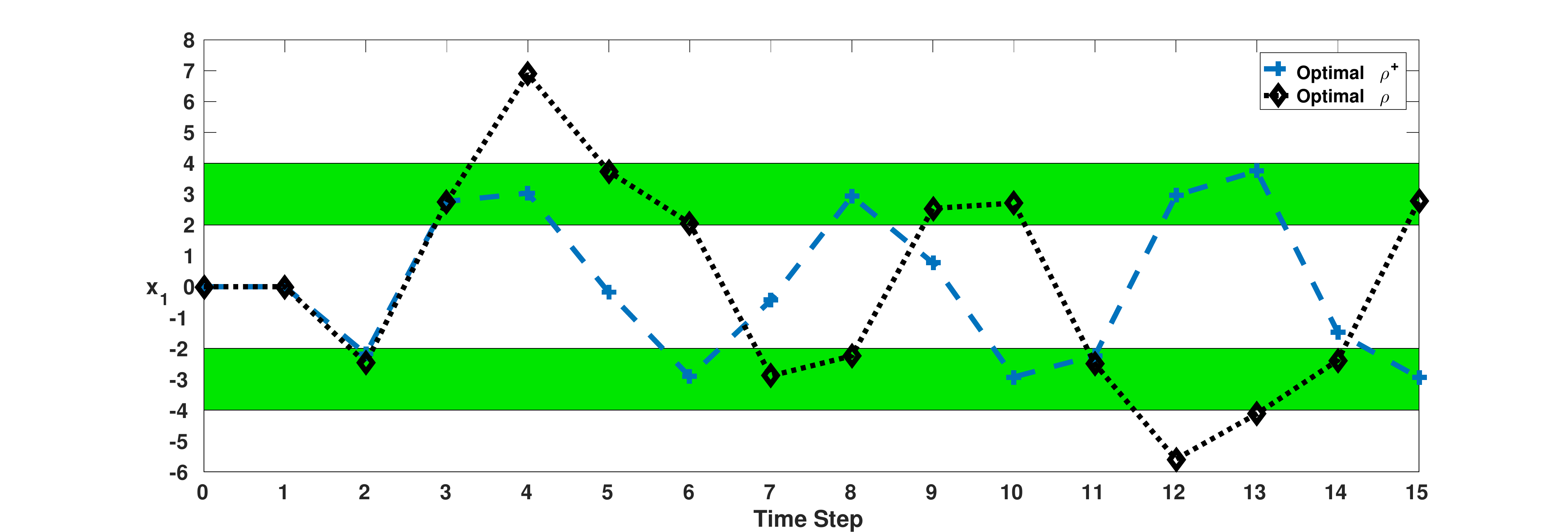}
\caption{Evolution of $x_1$ resulting from optimal control policies}
\label{fig:state2}
\end{figure}

Consider the system and specification of Example \ref{ex2} with the dynamics shown in \eqref{eq:ex2dynamics} and specification $\phi_2$ presented in \eqref{eq:ex2formula}.
We used Algorithm \ref{alg:optimization} in conjunction with the MPC framework as described in Section \ref{sec:mpc} to solve Example \ref{ex2} with optimal cumulative robustness. 
The computation took 6.73 seconds.
Additionally, we computed the control policy derived from only optimizing the traditional robustness $\tilde\rho$.

The corresponding evolution of $x_1$ according to the system dynamics \eqref{eq:ex2dynamics} for both cases is presented in Fig. \ref{fig:state2}. 
According to \eqref{eq:ex2formula}, the goal is for $x_1$ to periodically visit both of the green regions in this figure, always visiting each region within at most 4 time steps in the future.
Fig. \ref{fig:state2} shows that both control policies satisfy this requirement.
However, the trajectory that results from optimizing $\tilde\rho^+$ (dashed blue) tends to stay in the green regions as long as possible.
On the other hand, by optimizing the traditional STL robustness $\tilde\rho$, we are only ensuring that the green regions are visited once every $4$ time steps, and do not have any control over the duration of satisfaction. Fig. \ref{fig:state2} shows that the dotted black trajectory visits the green regions, but does not stay in them.In fact, it goes beyond the green regions three times.

\subsection{MPC for a Noisy Linear System}
We have demonstrated in previous case studies that cumulative robustness has a more desirable performance than conventional STL robustness. 
Now, we discuss the possibility that cumulative robustness can result in more robust control policies once noise is introduced to the system. 
Intuitively, since cumulative robustness adds up STL robustness over time, it may provide the added benefit of canceling out some of the perturbations caused by environmental noise over time.
We investigate this phenomenon in this case study.

Assume that we introduce disturbance to the system considered in the previous example:
\vspace{-5pt}
\begin{equation}
\label{eq:ex3noise}
x[k+1]=\begin{bmatrix}1&0.5\\0&0.8\end{bmatrix}x[k]+\begin{bmatrix}0\\1\end{bmatrix}u[k]+w[k],
\end{equation}
where $w[k]\in\mathbb{R}^2$ and $w_i[k]\thicksim\mathcal{N}(0,0.1)$ represents system disturbances with a normal distribution of $0$ mean and $0.1$ variance.
We consider the same specification as before \eqref{eq:ex2formula}.

We employed statistical model checking~\cite{zuliani2010bayesian} to investigate how the control policies that were synthesized in Section \ref{sec:case2} perform if system disturbances are considered.
The Bayesian estimation algorithm from~\cite{zuliani2010bayesian} was implemented, which computes the probability that a stochastic system satisfies a given specification, given a pre-determined margin of error and confidence level (i.e., the likelihood that the actual probability of satisfaction fall within the margin of error from the reported probability of satisfaction by the algorithm). The results are presented in Table \ref{tab:smc}.
\begin{table}
  \caption{Probability that a noisy linear system \eqref{eq:ex3noise} satisfies the STL specification of \eqref{eq:ex2formula}.}
  \label{tab:smc}
\begin{center}
\footnotesize
\begin{tabular}{ c  c  c  c  c }
\toprule
\shortstack{Objective \\ Function} & \shortstack{Probability of \\ Satisfaction} & \shortstack{Simulation \\ Sample Size} & \shortstack{Margin of \\ Error} & \shortstack{Confidence \\ Level} \\
\midrule
$\tilde\rho^+$ & 44.9\% & 328 & 1\% & 95\% \\
$\tilde\rho$ & 10.5\% & 273 & 1\% & 95\% \\
\bottomrule
\end{tabular}
\end{center}
\end{table}
\normalsize

\begin{figure}
\centering
\includegraphics[width=0.8\columnwidth]{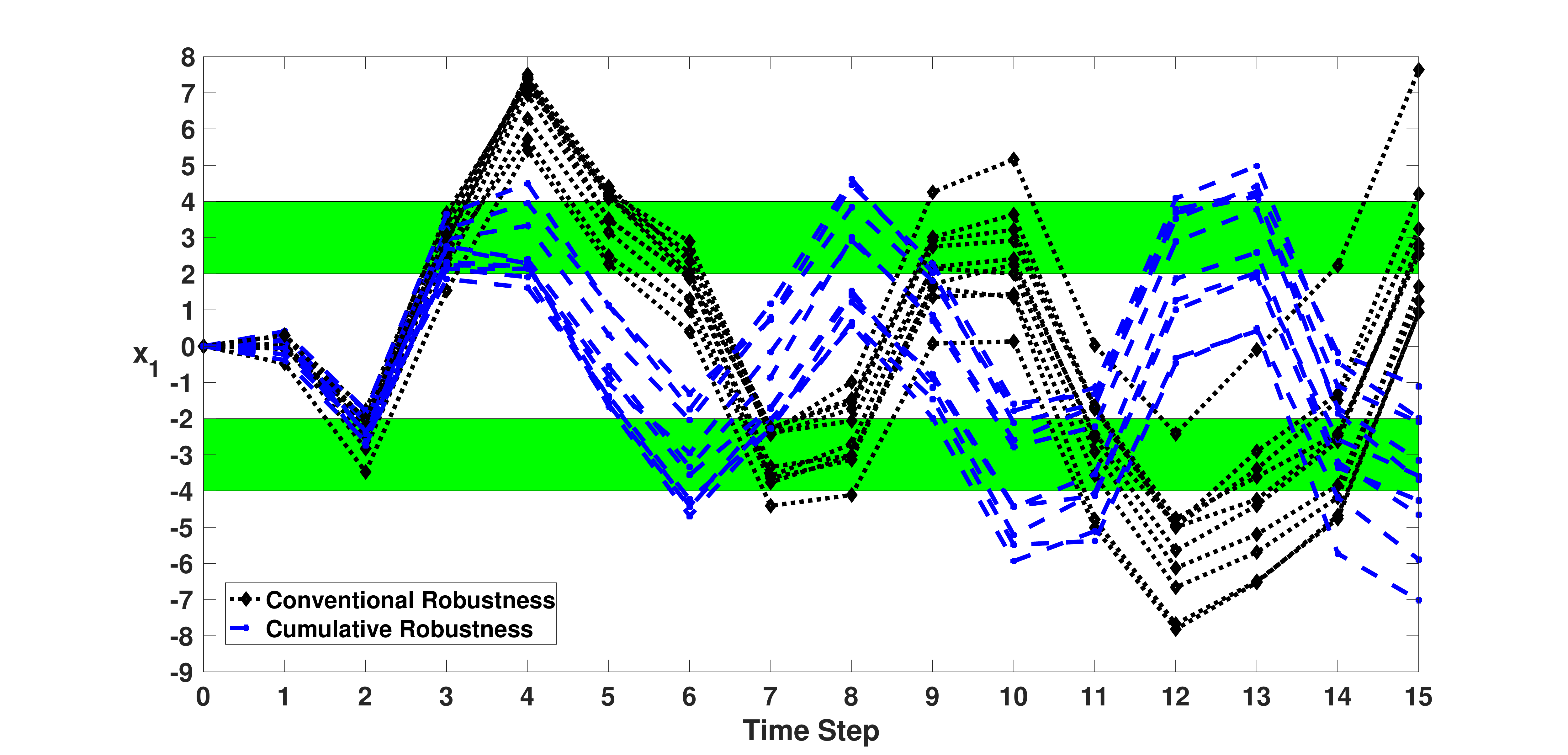}
\caption{The performance of control policies from Section \ref{sec:case2} if noise is added to the system dynamics \eqref{eq:ex3noise}.}
\label{fig:noise}
\end{figure}

Fig. \ref{fig:noise} illustrates 20 sample trajectories from the system of \eqref{eq:ex3noise} resulting from the control policies synthesized in the previous example (Section \ref{sec:case2}). 
Recall that the specification was for $x_1$ to reach the green regions periodically within $4$ time steps. 
Indeed, the trajectories derived from cumulative robustness (blue) are more resistant to disturbances and follow the specification much more consistently.

\begin{remark}
The intention in this case study was merely to demonstrate that cumulative robustness has the potential to improve controller performance under uncertainty and noise. We plan to study a provably robust MPC framework for this robustness function in the future.
\end{remark}


\bibliographystyle{IEEEtran}
\bibliography{references}

\end{document}